\documentclass[11pt]{article}
\usepackage{geometry}                
\usepackage{graphicx}
\usepackage{amsmath}
\usepackage{amssymb}
\usepackage{hyperref}

\newcommand{\abs}[1]{\left\lvert #1 \right\rvert}

\title{Rebuttal of `Note on ``Vacuum stability of a general scalar potential of a few fields'''}
\author{Kristjan Kannike
\\
\footnotesize
 National Institute of Chemical Physics and Biophysics, R\"{a}vala 10, 10143 Estonia
}

\begin{document}
\maketitle

\begin{abstract}
We show that the recent `Note on ``Vacuum stability of a general scalar potential of a few fields''' [arXiv:2401.13863] erroneously misses the possibility that the Higgs portal term may have a different sign for different values of the two singlet fields. Due to this mistake, the derived vacuum stability conditions are sufficient, but not necessary.
\end{abstract}

Our paper \cite{Kannike:2016fmd} considered the vacuum stability or bounded-from-below conditions of the general scalar potential of the Higgs boson $H$ and two real scalars $\phi_{1}$ and $\phi_{2}$. In order to find vacuum stability conditions, it is enough to consider the positivity of the quartic part of the potential, given by
\begin{equation}
  V(\phi_{1}, \phi_{2}, \abs{H}^{2}) = \lambda_{H} \abs{H}^{4} + M^{2}(\phi_{1}, \phi_{2}) \abs{H}^{2} + V(\phi_{1}, \phi_{2}), 
\end{equation}
where
\begin{align}
  M^{2}(\phi_{1}, \phi_{2}) &= \lambda_{H20} \phi_{1}^{2} + \lambda_{H11} \phi_{1} \phi_{2} + \lambda_{H02} \phi_{2}^{2},
  \\
  V(\phi_{1}, \phi_{2}) &= \lambda_{40} \phi_{1}^{4} + \lambda_{31} \phi_{1}^{3} \phi_{2} + \lambda_{22} \phi_{1}^{2} \phi_{2}^{2} + \lambda_{13} \phi_{1} \phi_{2}^{3} + \lambda_{04} \phi_{2}^{4}.
\label{eq:V:general:2:fields}
\end{align}
Considering the potential $V(\phi_{1}, \phi_{2}, \abs{H}^{2})$ as a quadratic polynomial in $\abs{H}^{2}$, we see that it is positive for all field values if and only if
\begin{align}
  \lambda_{H} &> 0, \quad V(\phi_{1}, \phi_{2}) > 0, \\
  M^{2}(\phi_{1}, \phi_{2}) &\geq 0 \; \lor \;4  \lambda_{H} V(\phi_{1}, \phi_{2}) - M^{4}(\phi_{1}, \phi_{2}) > 0.
  \label{eq:2:field:cond}
\end{align}
The coefficient matrix of the quadratic form $M^{2}(\phi_{1}, \phi_{2})$ can be written as
\begin{equation}
  \mathbf{M}^{2} = 
  \begin{pmatrix}
    \lambda_{H20} & \frac{1}{2} \lambda_{H11} \\
    \frac{1}{2} \lambda_{H11} & \lambda_{H02}
  \end{pmatrix}.
\end{equation}
The mistake of the recent `Note on ``Vacuum stability of a general scalar potential of a few fields'''~\cite{Song:2024yio} is  to not realise that the coefficient matrix $\mathbf{M}^{2}$ can be not only positive-definite or negative-definite, but also \emph{indefinite}. If $\mathbf{M}^{2}$ is positive-semidefinite, then $M^{2}(\phi_{1}, \phi_{2}) \geq 0$ holds for all values of $\phi_{1}$ and $\phi_{2}$. If $\mathbf{M}^{2}$ is negative-definite, then the second condition in Eq.~\eqref{eq:2:field:cond} must hold for all values of $\phi_{1}$ and $\phi_{2}$. For an indefinite $\mathbf{M}^{2}$, which has one negative and one positive eigenvalue, we have in a region of $\phi_{1}$ and $\phi_{2}$ that  $M^{2}(\phi_{1}, \phi_{2}) \geq 0$, while in another region with $M^{2}(\phi_{1}, \phi_{2}) < 0$ we must demand that the second, stronger condition in Eq.~\eqref{eq:2:field:cond} be satisfied.

The determinant of an indefinite $\mathbf{M}^{2}$ is negative, i.e. $4 \lambda_{H20} \lambda_{H02} - \lambda_{H11}^{2} < 0$. Unlike our work \cite{Kannike:2016fmd}, Ref.~\cite{Song:2024yio} erroneously does not take into account this possibility. Therefore the presented vacuum stability conditions, Eq.~(9) of Ref.~\cite{Song:2024yio}, are sufficient, but not necessary.

\section*{Acknowledgments}

This work was supported by the Estonian Research Council grants PRG803, RVTT3 and RVTT7, and by the EU through the European Regional Development Fund CoE program TK202 ``Fundamental Universe''.

\bibliographystyle{JHEP}
\bibliography{rebuttal}

\end{document}